\documentclass[conference]{IEEEtran}
\IEEEoverridecommandlockouts
\usepackage{cite}
\usepackage{amsmath,amssymb,amsfonts}
\usepackage{algorithmic}
\usepackage{graphicx}
\usepackage{textcomp}
\usepackage{xcolor}
\usepackage{enumerate}
\def\BibTeX{{\rm B\kern-.05em{\sc i\kern-.025em b}\kern-.08em
    T\kern-.1667em\lower.7ex\hbox{E}\kern-.125emX}}
\begin{document}

\title{Augmenting Channel Simulator and Semi- Supervised Learning for Efficient Indoor Positioning\\
{}
}

\author{\IEEEauthorblockN{Yupeng Li\textsuperscript{\dag}, Xinyu Ning\textsuperscript{\ddag}, Shijian Gao\textsuperscript{*}, Yitong Liu\textsuperscript{\ddag}, Zhi Sun\textsuperscript{\$}, Qixing Wang\textsuperscript{\dag}, Jiangzhou Wang\textsuperscript{\dag}} 
 	\IEEEauthorblockA{
\textsuperscript{\dag}China Mobile Research Institute, Beijing, China \\
\textsuperscript{\ddag}Beijing University of Posts and Telecommunications, Beijing, China\\
\textsuperscript{*}The Hong Kong University of Science and Technology (Guangzhou), China\\
\textsuperscript{\$}Tsinghua University, Beijing, China \\
  }
  }

\maketitle

\begin{abstract}
This work aims to tackle the labor-intensive and resource-consuming task of indoor positioning by proposing an efficient approach. The proposed approach involves the introduction of a semi-supervised learning (SSL) with a biased teacher (SSLB) algorithm, which effectively utilizes both labeled and unlabeled channel data. To reduce measurement expenses, unlabeled data is generated using an updated channel simulator (UCHS), and then weighted by adaptive confidence values to simplify the tuning of hyperparameters. Simulation results demonstrate that the proposed strategy achieves superior performance while minimizing measurement overhead and training expense compared to existing benchmarks, offering a valuable and practical solution for indoor positioning.



\end{abstract}


\section{Introduction}
The 3rd Generation Partnership Project (3GPP) has successfully conducted the fifth generation (5G) Advanced phase in its Release 18\cite{R18}. As one of the most significant potential advancements to increase automation and efficiency of the industrial domain, the indoor positioning has attracted widespread attention. However, the indoor factory scenario presents challenges, including a high volume of obstructions leading to numerous non-line-of-sight (NLOS) scenarios. These conditions often lead to significant performance degradation for the traditional positioning algorithms \cite{geome1,geome4}. 


To address the issue of the limitations and difficulties of traditional positioning methods, a notable point is to integrate deep learning (DL) into indoor positioning systems, specifically fingerprint-based positioning. 
The basic idea of the fingerprint positioning is to create a mapping function between the channel measurements and the position coordinates, creating a main wireless radio environment among the transceivers \cite{fingerprint2}. Then, the user equipment (UE) can obtain its position by the measurements. The measurements of fingerprint positioning in the cellular systems are mainly channel impulse response (CIR), power delay profile (PDP), and so on \cite{TSG_RAN}.
However, DL-based indoor positioning faces the challenge of insufficient data available for training the model \cite{insufficient}. Especially, in the flexible and dynamic communication systems, obtaining sufficient measurements and labeling them is a labor-intensive and resource-consuming task. 
A powerful approach termed semi-supervised learning (SSL) was proposed, which involved using small number of labeled data combined with a large amount of unlabeled data to train the model \cite{ssl,ssl1,ssl2}. 
Typical SSL \cite{pseudo} may lead to serious performance degradation due to errors in the pseudo-labels of the unlabeled data.
To address this issue, a smaller weighting coefficient was added to the unlabeled data during model training \cite{semi_and_gan}.
The weighting coefficient is a hyperparameter, which requires meticulous manual tuning.
In \cite{fixmatch}, the concept of confidence for the pseudo-label of the unlabeled data was proposed, which helps to select reliable unlabeled data in classification tasks. However, there is a lack of a method to define a label confidence in regression tasks like positioning. Furthermore, in an indoor scenario, obtaining the unlabeled data is also a challenge that requires extra measurement overhead.

To address the aforementioned challenges, a DL-based positioning algorithm with reduced overhead is proposed. To lower the measurement overhead, the unlabeled data is generated by an updated channel simulator according to the stochastic parameters of the labeled data. This approach reduces the measurement expenses of the unlabeled data. Additionally, to capture the features of both the labeled and unlabeled channel data, a semi-supervised learning with a biased teacher is proposed. The main contributions of this paper are summarized as follows:

\begin{itemize}
\item An updated channel simulator is utilized to generate unlabeled data, thereby reducing the measurement expenses associated with acquiring such data. Initially, labeled data collected within an indoor factory environment is utilized, and the primary channel parameters are analyzed to extract the channel stochastic parameters of the measurement environment. Subsequently, the typical TR 38.901 \cite{901} channel simulator is updated with these stochastic parameters, resulting in the creation of an updated channel simulator (UCHS). The UCHS is then employed to simulate the unlabeled channel data, which closely resembles the labeled data of the measurement environment.

\item To efficiently capture the mapping function between the CIR and the position coordinate while minimizing training expenses, a SSL with a biased teacher (SSLB) is introduced. The SSLB approach involves weighting the unlabeled data based on the confidence, which is derived from the distance bias predicted by the teacher model. This methodology assigns higher confidence to more reliable unlabeled data. Notably, this work represents the first instance of introducing confidence into regression tasks, such as indoor positioning, resulting in a significant reduction in training expenses.
\item To obtain the confidence of the unlabeled data, a novel SSL structure is designed, termed as SSLB. This structure incorporates an additional reference signal and a position bias, added separately to the input and the output of the traditional SSL model. As a result, SSLB can extract additional bias information from the labeled dataset, which can then be transformed into the confidence of the unlabeled data. 


\end{itemize}

\textit{Notation:} Vectors and matrices are denoted by boldface. $||\cdot||$ expresses the 2-norm of a vector. $||\cdot||_{\text{F}}$ is the Frobenius norm of a matrix. For unlabeled data, we use a superscript $^*$, while labeled data remains unmarked.

\section{Problem description}



\subsection{Adapting semi-supervised learning for indoor positioning}
The dataset of traditional SSL consists of two parts: the labeled dataset and the unlabeled dataset. The input is the CIR and the output is the predicted position.
The labeled dataset is represented as $\mathfrak{D}(\textit{\textbf{H}}, \textit{\textbf{l}})$, where $\textit{\textbf{H}}$ is the labeled CIR and $\textit{\textbf{l}}$ is its corresponding position coordinate. The unlabeled dataset is represented as $\mathfrak{E}(\textit{\textbf{H}}^*)$, containing only the CIR.

\begin{figure}[!t]
\centering
\includegraphics[width=0.85\linewidth]{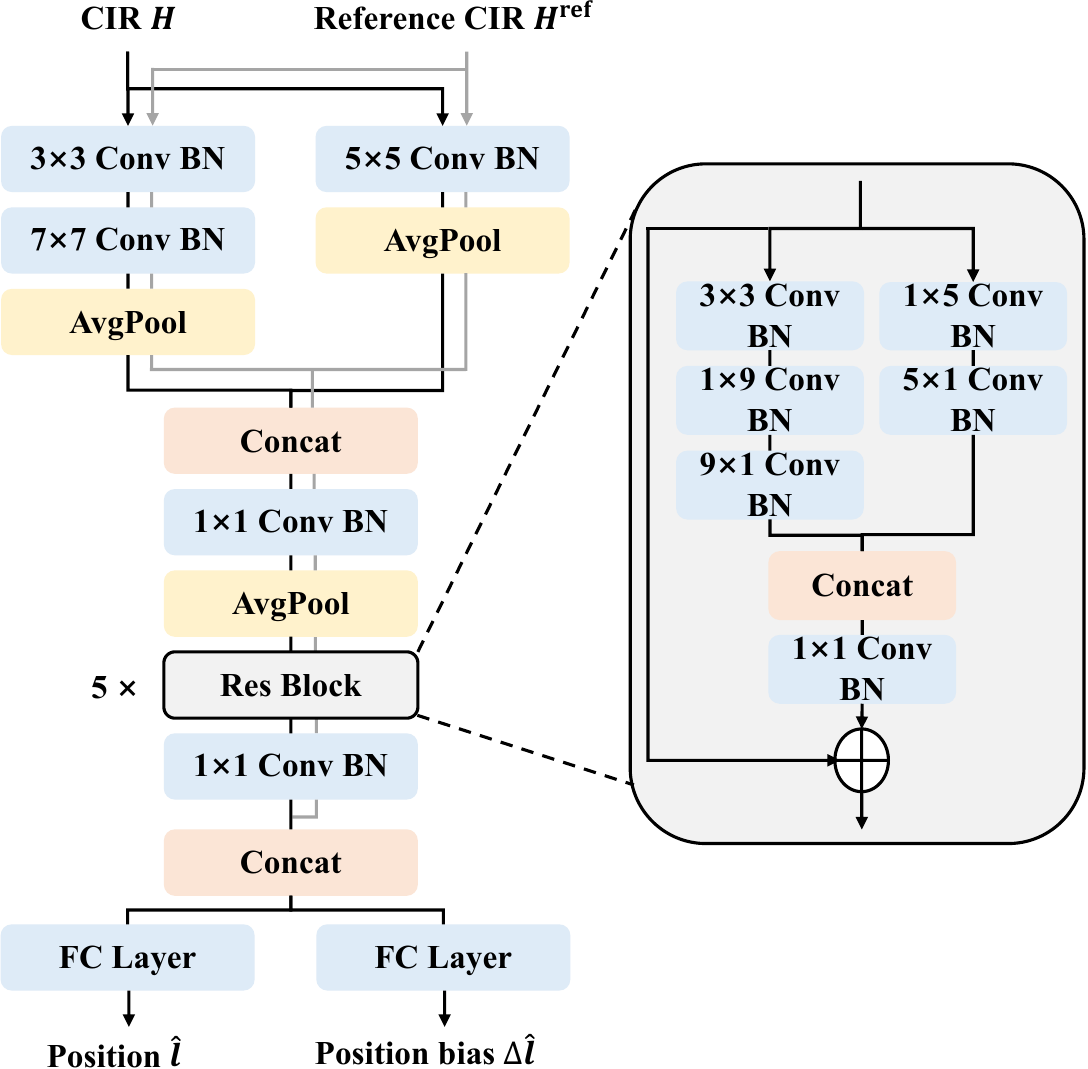}
\caption{The DL model for the indoor positioning.}\label{fig:model}
\end{figure}

The SSL model is divided into a teacher model $\Theta^\text{t}$ and a student model $\Theta^\text{s}$. Firstly, the teacher model is trained with labeled data, and then pseudo-labels of unlabeled data are generated using the trained teacher model. Finally, the student model is trained with both the labeled data and the unlabeled data. In this paper, the loss function is defined as the Euclidean distance between the predicted position and the real position as shown in (\ref{equ:euclidean_loss}). First, train the teacher model using this loss on the labeled dataset $\mathfrak{D}(\textit{\textbf{H}}, \textit{\textbf{l}})$.
\begin{equation}
L = \frac{1}{N} \sum^N_{i=1} ||\hat{\textit{\textbf{l}}}_{i} - \textit{\textbf{l}}_i||\label{equ:euclidean_loss}.
\end{equation}
Then, the label of dataset $\mathfrak{E}(\textit{\textbf{H}}^*)$ is generated with the trained teacher model
\begin{equation}
\textit{\textbf{l}}^* = f(\textit{\textbf{H}}^*, \Theta^\text{t}),\label{equ:pesudo_label}
\end{equation}
where $\textit{\textbf{l}}^*$ is the position coordinate of $\textit{\textbf{H}}^*$. $f(\cdot)$ represents the mapping function between the channel measurements and position coordinates. In this way, the dataset $\mathfrak{E}(\textit{\textbf{H}}^*)$ is completed to $\mathfrak{E}(\textit{\textbf{H}}^*, \textit{\textbf{l}}^*)$.
Finally, the student model is trained with the dataset of $\mathfrak{D}(\textit{\textbf{H}}, \textit{\textbf{l}})$ and $\mathfrak{E}(\textit{\textbf{H}}^*, \textit{\textbf{l}}^*)$, applying weights to pseudo-labeled data. The loss function is defined as follows:
\begin{equation}
\begin{aligned}
\hspace{-0.16cm}L^{\text{w}} \hspace{-0.1cm}&=\hspace{-0.1cm} \frac{1}{N\hspace{-0.1cm}+\hspace{-0.1cm}N^*}\left(\sum^{N}_{i=1}\hspace{-0.1cm}\sqrt{w||\hat{\textit{\textbf{l}}}_i\hspace{-0.1cm}-\hspace{-0.1cm}\textit{\textbf{l}}_{i}||^2} +  \sum^{N^*}_{j=1}\hspace{-0.1cm}\sqrt{w^*||\hat{\textit{\textbf{l}}}^*_j\hspace{-0.1cm}-\hspace{-0.1cm}\textit{\textbf{l}}^*_{j}||^2} \right),\label{equ:weight_loss}
\end{aligned}
\end{equation}
where $L^{\text{w}}$ is the loss of weighted semi-supervised algorithm, $N$ and $N^*$ are the numbers of labeled data and unlabeled data, and $w$ and $w^*$ are the corresponding weights of labeled data and unlabeled data. Detailed descriptions can be referred to \cite{pseudo}\cite{semi_and_gan}. it is worth mentioning that all models use the same DL model, illustrated in Fig. \ref{fig:model}.

\section{Channel feature aided data augmentation}


\subsection{Wireless channel feature extracting method}\label{modeling}

In this section, some specific methods for extracting channel parameters will be described.

\subsubsection{Angle spread}
The angle spread (AS) is used to characterize the dispersion features of the multipaths in the spatial domain. Initially, a two-dimensional Discrete Fourier transform (DFT) is performed over the channel matrix $\textit{\textbf{H}}(t)$ over the port domain, resulting in the angle domain channel matrix $\textit{\textbf{H}}_{\text{ang}}$ \cite{901} as follows:
\begin{equation}
\textit{\textbf{H}}_{\text{ang}} = \textit{\textbf{U}}^T \textit{\textbf{H}}(t) \textit{\textbf{V}},
\end{equation}
where $\textit{\textbf{U}} = [\textit{\textbf{u}}_1^T; ...; \textit{\textbf{u}}_{N_t}^T]^T$ is the transmit DFT matrix, and $\textit{\textbf{V}} = [\textit{\textbf{v}}_1, ... ,\textit{\textbf{v}}_{N_r}]$ is the receiver DFT matrix\cite{ruisi}.

Then, the receiving and transmitting AS are illustrated in the spatial domain. The angle spread value can be obtained as follows:
\begin{equation}
\sigma_{AS} = \sqrt{\frac{ \sum^{M}_{m=1} {\theta_{m}^2P_m} }{\sum^{M}_{m=1} {P_m}}}
,\end{equation}
where $P_m$ is the power of the $m$th multipath, $\theta_m$ is the angle of the $m$th multipath.

\subsubsection{Delay spread}
The delay spread (DS) can reflect the richness of the channel multipaths. And the DS \cite{extrpara} can be obtained as follows:
\begin{equation}
\sigma_{DS} = \sqrt{\frac{ \sum^{M}_{m=1} {(\tau_{m} - \tau_{\text{mean}})^2 P_m} }{\sum^{M}_{m=1} {P_m}}},
\end{equation}
where $\tau_{mean} = \sum^{M}_{m=1} ({\tau_m}P_{m})/{\sum^{M}_{m=1} (P_m)}$ is the mean delay.

Due to the space limit, only the above two unique channel statistical parameters are delineated here. Other channel statistics parameters can refer to \cite{ruisi}.

\subsection{Unlabeled data generation via UCHS}

The TR 38.901 \cite{901}, considered as the state-of-the-art 3GPP channel model, is crucial for generating channel simulation data and evaluating physical layer techniques. It illustrates the stochastic distribution characteristics of channel multipaths, serving as a benchmark in a general scenario. However, in a specific location, the channel parameters in TR 38.901 may not be precise. Regrettably, in an AI-enabled indoor factory, the training data needs to reflect the actual propagation environment as accurately as possible.

To address the aforementioned issue, we propose a data augmentation method based on the channel simulator, which can simulate the unlabeled channel data corresponding to the specific area. This approach significantly reduces the collection overhead of unlabeled data. 
The generation process of unlabeled data is shown as follows:
\begin{itemize}
\item \textbf{Step 1}: the labeled channel data is collected by the UEs distributed in a specific indoor factory. Since labeled data is essential for training the teacher model, there is no extra overhead of data collection at this
stage.
\item \textbf{Step 2}: upload the labeled data from the UE to the BS.
\item \textbf{Step 3}: the primary statistical parameters of the channel data are extracted, as described in Section \ref{modeling}, effectively depicting the wireless propagation environment.
\item \textbf{Step 4}: these statistical parameters are used to update the parameters of
typical TR 38.901 channel model, resulting in the UCHS.
\item \textbf{Step 5}: the UCHS is used to simulate unlabeled channel data similar to the labeled data of the measurement environment.
\end{itemize}

\section{Semi-Supervised Pseudo-Labeling with a Biased Teacher}
\label{sec:bias_teacher}
\begin{figure*}[!t]
\centering
\includegraphics[width=0.7\linewidth]{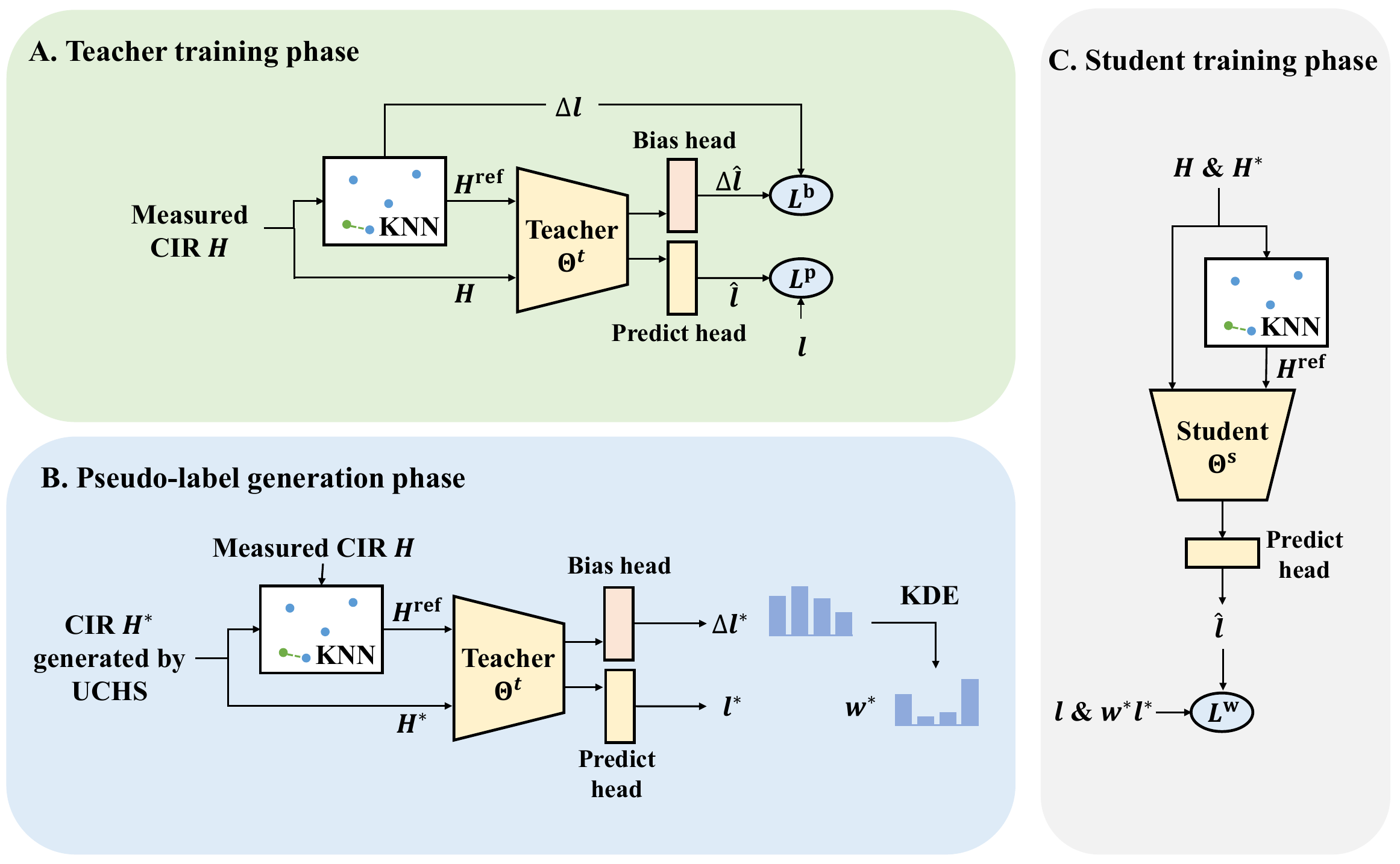}
\caption{The pipeline of semi-supervised learning with a biased teacher.}\label{fig:pipeline}
\end{figure*}

Compared to traditional SSLs, our proposed method assigns different loss weights to the pseudo-labels of the unlabeled data based on the confidence scores. By doing so, pseudo-labels with low confidence are assigned small weights during the training stage of the student model, thereby mitigating the impact of pseudo-labels with large offset on the performance of the student model.


The proposed method pipeline, as illustrated in Fig. \ref{fig:pipeline}, consists of three primary stages: the bias teacher training stage, the confidence-based pseudo-labeling stage, and the weighted loss student training stage:
\begin{itemize}
\item \textbf{Stage 1}: an additional reference signal and a position bias are separately added to the input and output of the traditional SSL model. This addition helps extract additional information from the labeled dataset. Specifically, the position bias represents the distance between the input signal and its nearest signal, referred to as the reference signal. This allows the teacher model to not only determine the mapping function between the input signal and the position coordinates but also to assess the similarity between input signals using the distance bias.
\item \textbf{Stage 2}: the trained teacher model is utilized to generate pseudo-labels and the position bias of the unlabeled data. The position bias can then be used to define the confidence scores of the unlabeled data through a kernel density estimation (KDE) \cite{kde} function.
\item \textbf{Stage 3}: a weighted loss strategy is employed, wherein different loss weights are assigned to different pseudo-labels in the loss function to train the student model. This approach aims to mitigate the impact of pseudo-labels with larger errors on the performance of the student model.
\end{itemize}
\subsection{The biased teacher training}
\label{sec:bias_teacher_training}
In the context of object detection, models not only output the bounding boxes of the detected objects but also provide confidence scores for these detections. Inspired by this, a biased teacher model with two prediction heads is proposed. One prediction head, called the 'predict head'', is responsible for predicting the position coordinates $\hat{\textit{\textbf{l}}}$ corresponding to the input CIR. The other prediction head, named as the ''biased head'' is designed to predict the bias $\Delta\hat{\textit{\textbf{l}}}$ of the reference signal relative to the input CIR position.


The first step involves using the k-nearest neighbors (KNN)\cite{knn} algorithm to search the reference signal $\textit{\textbf{H}}^{\text{ref}}_i$ in labeled dataset $\mathfrak{D}(\textit{\textbf{H}}, \textit{\textbf{l}})$ that is closest to the input CIR $\textit{\textbf{H}}_i$, thereby obtaining the reference signal data pair $(\textit{\textbf{H}}^{\text{ref}}_i, \textit{\textbf{l}}^{\text{ref}}_i)$, i.e.,

\begin{equation}
\textit{\textbf{H}}^{\text{ref}}_i = \textit{\textbf{H}}_j, \mathop{\arg\min}_{j \in (1, 2, ..., N)}  ||\textit{\textbf{H}}_i-\textit{\textbf{H}}_j||_{\text{F}}, i \neq j ,
\label{equ:knn}
\end{equation}
where $|| \cdot ||_{\text{F}}$ is Frobenius norm, and $i \neq j$ excludes itself as the nearest neighbor. 

The position bias is defined as:
\begin{equation}
\Delta \textit{\textbf{l}}_i \triangleq \textit{\textbf{l}}^{\text{ref}}_i - \textit{\textbf{l}}_i\label{equ:pos_bias}.
\end{equation}
The total loss function of the teacher model is defined as:
\begin{equation}
L = L^{\text{p}} + L^{\text{b}},\label{equ:loss}
\end{equation}
where $L^{\text{p}} = \frac{1}{N} \sum\limits_{i = 1}^N ||\hat{\textit{\textbf{l}}_i}-\textit{\textbf{l}}_i||$ and $L^{\text{b}} = \frac{1}{N} \sum\limits_{i = 1}^N ||\Delta \hat{\textit{\textbf{l}}_i}-\Delta \textit{\textbf{l}}_i||$ are the position prediction loss and the bias prediction loss, respectively.



\subsection{Confidence-based pseudo-labeling}
\label{sec:pseudo_phase}
This process involves using the teacher model to generate estimated positions $\textit{\textbf{l}}^*$ and position biases $\Delta \textit{\textbf{l}}^*$ for the unlabeled data, which can then be used for subsequent steps in the SSL framework.

\begin{figure}[htbp]
    \centering
    \begin{minipage}[b]{0.35\textwidth}
        \centering
        \includegraphics[width=\textwidth]{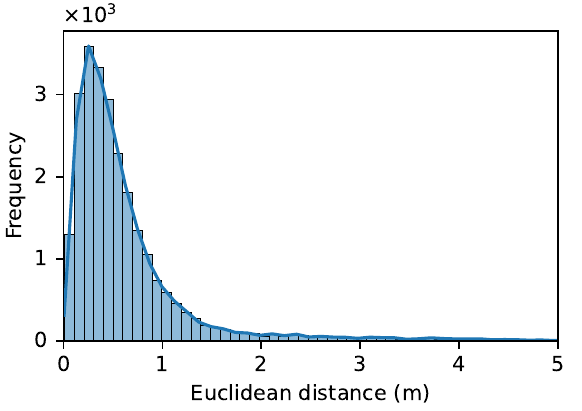}
        {\scriptsize(a) Distribution of predicted position bias}
        \label{fig:bias_distribution_a}
    \end{minipage}
    \hfill
    \begin{minipage}[b]{0.35\textwidth}
        \centering
        \includegraphics[width=\textwidth]{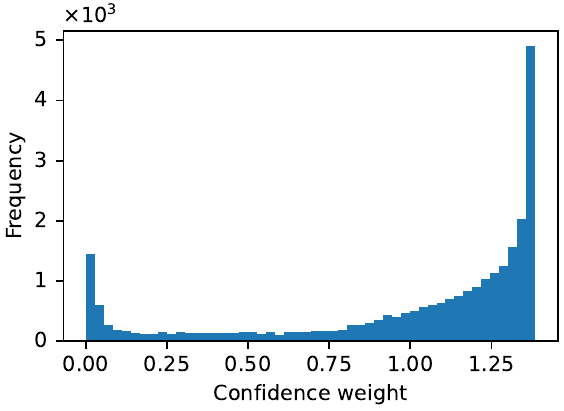}
        {\scriptsize(b) Distribution of confidence weight}
        \label{fig:bias_distribution_b}
    \end{minipage}
    \caption{Distribution of predicted position bias and confidence weight.}
    \label{fig:bias_distribution}
\end{figure}

For the unlabeled data, the confidence is determined by calculating the distribution of the position bias. This involves computing the position bias as $d=||\Delta \textit{\textbf{l}}^*||$. Subsequently, the distribution can be fitted using KDE as follows:
\begin{equation}
\hat{f}_h(d)=\frac{1}{Nh}\sum^N_{i=1}K\left( \frac{d-d_i}{h}\right)\label{equ:kde}
,\end{equation}
where $\hat{f}_h(d)$ is the KDE at point $d$. $N$ is the number of samples, $h$ is the bandwidth, $d_i$ represents each data point in the sample, $K(\cdot)$ represents the kernel function, and in this work the Gaussian kernel is used. Fig. \ref{fig:bias_distribution}(a) illustrates the distribution of the unlabeled data in our experiment.

The confidence is defined as:
\begin{equation}
w^*_i =\hat{f}_h(d_i)\label{equ:confidence_weight}.
\end{equation}
To ensure a fair comparison with the SSL, the obtained weights $w^*_i$ need to divide the mean value of all the $w^*$ to guarantee its mean value to be 1. This normalization process is important for comparison purposes. The distribution of $w^*$ is shown in Fig. \ref{fig:bias_distribution}(b).

\subsection{Weighted loss for student training}
The student model $\Theta^{\text{s}}$ is trained with the mixture dataset, which is constructed by the unlabeled dataset $\mathfrak{E}(\textit{\textbf{H}}^*,\textit{\textbf{l}}^*,w^*)$ 
combined with the labeled dataset $\mathfrak{D}(\textit{\textbf{H}}, \textit{\textbf{l}})$. 

For the student model, only the predicted coordinates $\hat{\textit{\textbf{l}}}$ are used to compute the loss, which is defined as
\begin{equation}
\begin{aligned}
\hspace{-0.16cm}L^{\text{w}} \hspace{-0.1cm}&=\hspace{-0.1cm} \frac{1}{N\hspace{-0.1cm}+\hspace{-0.1cm}N^*}\left( \sum^{N}_{i=1}\hspace{-0.1cm} \ ||\hat{\textit{\textbf{l}}_i}\hspace{-0.1cm}-\hspace{-0.1cm}\textit{\textbf{l}}_i|| + \sum^{N^*}_{j=1}\hspace{-0.1cm}\sqrt{w^*_i||\hat{\textit{\textbf{l}}}^*_j\hspace{-0.1cm}-\hspace{-0.1cm}\textit{\textbf{l}}^*_j||^2}\right),\label{equ:adapted_weight_loss}
\end{aligned}
\end{equation}
where $w^*_i$ is the weight of the pseudo-label, and the weight of labeled data is set as 1.


\section{Experiments and Results}
\subsection{Scenario deployment}
The dense clutter high BS (DH) sub-scenario is a priority for evaluation in 3GPP for AI-enabled indoor factory (InF) \cite{R18}. In our considered scenario, there are 18 BSs uniformly distributed within the 60m*120m space with a spacing of 20m and height of 8m. The central frequency is 3.5 GHz and the bandwidth is 100 MHz in our experiment. The clutter density is 40\% with the clutter height of 2m. The UEs are randomly generated within the space with a height of 1.5m.


\subsection{Experimental setup}
In the experiment, a total of 17200 labeled channel samples are collected in the measurement scenario, where 10000 samples are randomly selected to construct the training dataset and the residual 7200 samples are allocated to the testing dataset. Besides, the unlabeled channel data are generated using UCHS. 
All the models are implemented using PyTorch and the Adam optimizer. The hyperparameters are learning rate 5e-2, an input batch size of 256, 150 epochs, and a cosine annealing rate scheduler. The criteria for InF positioning is the positioning accuracy at a cumulative distribution function (CDF) of 90\% for users \cite{R18}.


\subsection{Data structure}
In the cellular positioning problem, UEs measure the time domain CIR $\textit{\textbf{H}}(t)$ from multiple BSs. For each position, the measured $\textit{\textbf{H}}(t)$ by each UE is in the form of $N_{BS}*N_{port}*N_{delay}*2$ dimensions, where $N_{BS}$, $N_{port}$, and $N_{delay}$ represent the number of BS, the number of BS antenna respectively, and the number of the time delay, and 2 is from the complex-domain to real and imaginary parts. In most industrial scenario, the effective delay is mainly distributed in the first 64 delay taps. Under the current setup, this results in an extremely high-dimensional tensor for each position in the form of $18*4*64*2$.

\subsection{Experimental results}
\subsubsection{Main results}
In this section, we carry out simulations to evaluate the performance of our proposed  SSL with position bias, which we call as “SSLB” scheme hereafter. We compare the proposed alrotihm with the following three benchmark schemes: 
\begin{itemize}
\item \textbf{a)} SL: traditional supervised learning, whose reference signal $\textit{\textbf{H}}^{\text{ref}}$ is represented by the input signal $\textit{\textbf{H}}$, with all the predicted position biases being zero. 
\item \textbf{b)} SLR: SL learning with reference signals. Its input signal contains the current CIR $\textit{\textbf{H}}$ and the reference signal $\textit{\textbf{H}}^{\text{ref}}$, and the output is $\hat{\textit{\textbf{l}}}$ and $\Delta \hat{\textit{\textbf{l}}}$. 
\item \textbf{c)} SSLR: SSL pseudo-label learning without a biased teacher. 
\end{itemize}The same DL model and training hyperparameters are used for the four schemes mentioned above. Our method aims to provide a convenient guideline for adaptively obtaining the weight of the unlabeled data using confidence. This is compatible with consistency-based pseudo-labeling SSL \cite{fixmatch,mean_teacher,flexmatch,usb}, and combining them may lead to better results, but it is beyond the scope of this paper.
 




\begin{figure}[]
\centering
\includegraphics[width=1\linewidth]{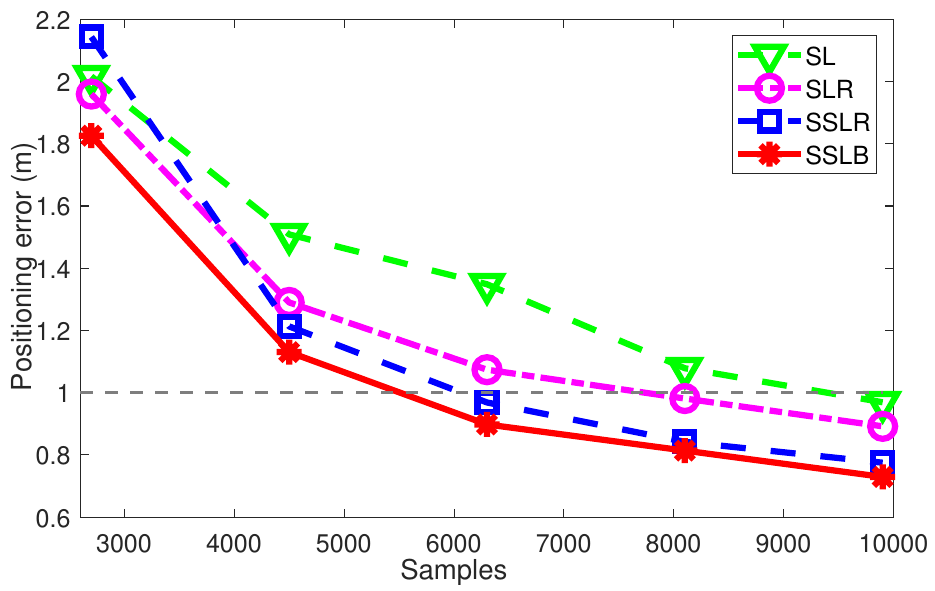}
\caption{Positioning accuracy comparison with different labeled samples.}\label{fig:main_result}
\end{figure}

Fig. \ref{fig:main_result} illustrates the positioning accuracy for the four schemes with different labeled samples. It is evident that the SSLB achieves the highest positioning precision. For instance, with less than 6300 samples, SSLB can achieve the precision of 0.897 m, while the SL requires 9900 samples to achieve a precision of 0.969 m. Therefore, with the help of UCHS, our proposed SSLB can reduce the data overhead by almost 40\% compared to SL. However, it is worth noting that the performance of the SSL pseudo-label method degrades at 2700 samples, due to the introduction of enormous inaccurate pseudo-labels by the teacher model, which deteriorates the performance of the student model. In contrast, SSLB achieves relatively stable performance, as it can automatically assign small weights to the unlabeled data with large offset.


\subsubsection{Ablation study}

\paragraph{The impact of weight on the test loss}

\begin{figure}[]
\centering
\includegraphics[width=0.96\linewidth]{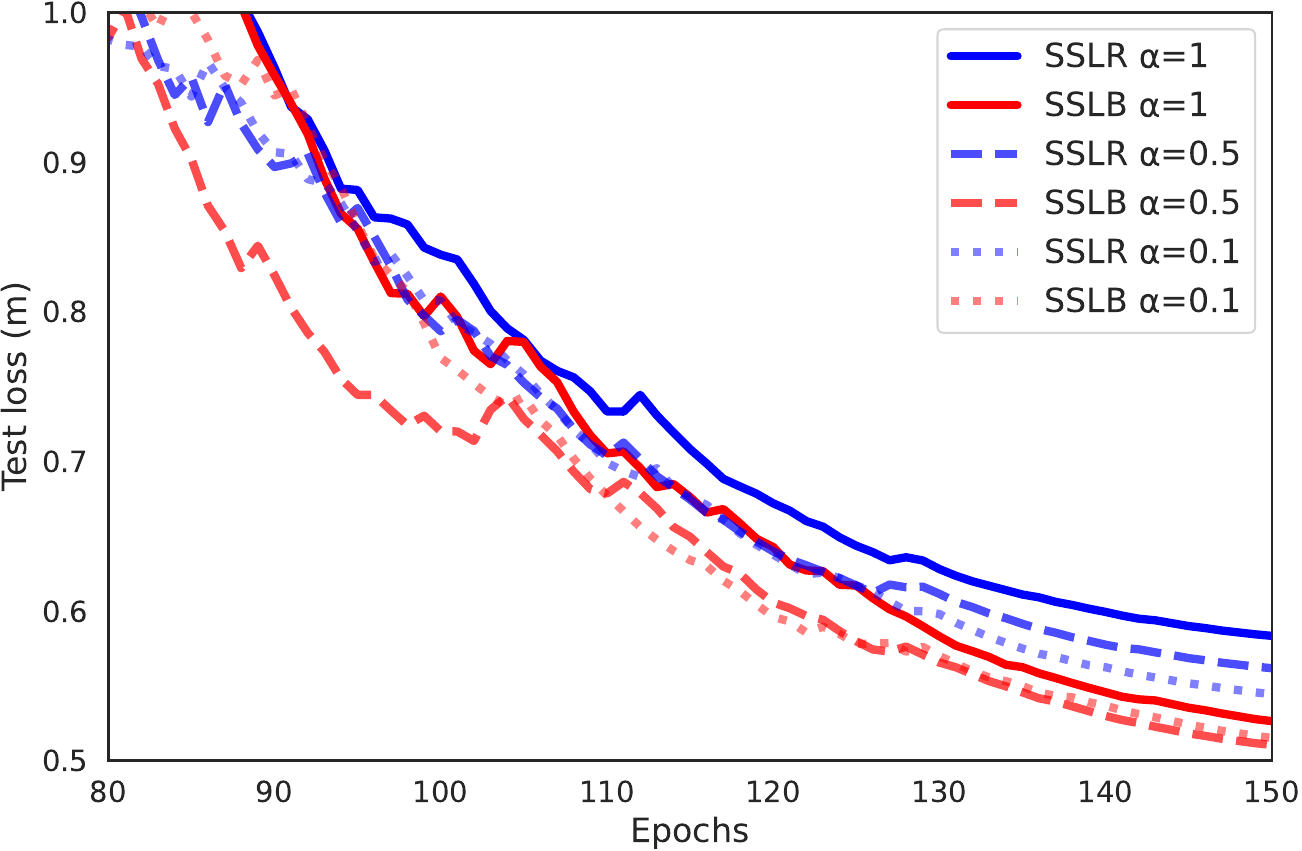}
\caption{Impact of weight magnitudes on the test loss.}\label{fig:ablation_weight}
\end{figure}

\begin{table}[]
\caption{Impact of weight on positioning accuracy}
\begin{center}
\begin{tabular}{p{1cm}p{1cm}p{1cm}p{1cm}}
\hline
 & $\alpha$ = 1 & $\alpha$ = 0.5 & $\alpha$ = 0.1\\
\hline
SSLR & 0.968 & 0.913 & 0.917\\
SSLB & \textbf{0.897} & \textbf{0.889} & \textbf{0.895}\\
\hline
\end{tabular}
\label{tab:ablation_weight}
\end{center}
\end{table}

In this section, we study the effects of weight magnitude $\alpha$ on the SSLR and our proposed SSLB. In the student model training stage, for both the SSLR and the SSLB, the unlabeled weights are scaled by the formula $w^{*'}_{i} = \alpha w^*_{i}$. Fig. \ref{fig:ablation_weight} compares the performance of SSLR and SSLB under different scaling coefficients. As the scaling coefficient decreases, the performance gap between the two schemes reduces, whereas the SSLB outperforms the SSLR in all cases. Table \ref{tab:ablation_weight} demonstrates the position accuracy at CDF of 90\% for users. With the decrease of scaling coefficients, the positioning precision of SSLR increases at first, and then decreases. However, the performance of KDE confidence based SSLB keeps relatively stable. Thus, our proposed scheme does not manually select the weights hyperparameters $\alpha$ of the unlabeled data and dynamically assigns different weights to the pseudo-labels. Therefore, the training expense is dramatically reduced.

\paragraph{Impact of confidence calibration}
To verify the effectiveness of the KDE confidence based method for transforming the position bias to confidence, we compare it with a linear transformation method. The linear transformation method directly considers position bias as confidence, assuming that small bias corresponds to high confidence levels. Firstly, the Euclidean distance of the position bias is computed. Then,  this data is normalized within the range of 0 to 1, reversed by subtracting from 1, and finally divided by the mean. The comparison results between the linear method and KDE are presented in Table \ref{tab:confidence_calibration}.

\begin{table}[]
\caption{Impact of confidence on positioning accuracy}
\begin{center}
\begin{tabular}{p{3cm}p{0.6cm}p{0.6cm}p{0.6cm}p{0.6cm}p{0.6cm}}
\hline
\textbf{Sample Number} & 2700 & 4500 & 6300 & 8100 & 9900\\
\hline
SSL w/o confidence & 2.143 & 1.213 & 0.968 & 0.842 & 0.775\\
SSL w/ linear confidence & 2.043 & 1.191 & 0.930 & 0.853 & 0.764\\
SSL w/ KDE confidence & \textbf{1.825} & \textbf{1.130} & \textbf{0.897} & \textbf{0.813} & \textbf{0.729}\\
\hline
\end{tabular}
\label{tab:confidence_calibration}
\end{center}
\end{table}

\section{Conclusion}
This paper has investigated the issue of indoor positioning with reduced expenses. On one hand, the the UCHS was used to generate unlabeled data and reduce measurement overheads. On the other hand, we proposed the SSLB, which adaptively weighted the unlabeled data based on confidence to differentiate the quality of the data. This approach effectively reduced the expenses associated with tuning the hyperparameters of the weights. Simulation results demonstrated that our proposed scheme exhibited superior performance, reducing the data measurement overhead by almost 40\% compared to benchmark schemes. Furthermore, ablation study confirmed that our proposed scheme can automatically assign different weights to unlabeled data, significantly reducing training expenses.

\vspace{12pt}

\end{document}